\documentclass[twocolumn]{aastex631}

\shortauthors{Glidden et al.}
\shorttitle{Isotopologues}

\graphicspath{{./}{figures/}}

\begin{document}

\title{Can Carbon Fractionation Provide Evidence for Aerial Biospheres in the Atmospheres of Temperate Sub-Neptunes?}

\correspondingauthor{Ana Glidden}
\email{aglidden@mit.edu}

\author[0000-0002-5322-2315]{Ana Glidden}
\affiliation{Department of Earth, Atmospheric and Planetary Sciences, Massachusetts Institute of Technology, Cambridge, MA 02139, USA}
\affiliation{Kavli Institute for Astrophysics and Space Research, Massachusetts Institute of Technology, Cambridge, MA 02139, USA}

\author[0000-0002-6892-6948]{Sara Seager}
\affiliation{Department of Earth, Atmospheric and Planetary Sciences, Massachusetts Institute of Technology, Cambridge, MA 02139, USA}
\affiliation{Department of Physics and Kavli Institute for Astrophysics and Space Research, Massachusetts Institute of Technology, Cambridge, MA 02139, USA}
\affiliation{Department of Aeronautics and Astronautics, Massachusetts Institute of Technology, Cambridge, MA 02139, USA}

\author[0000-0001-5732-8531]{Jingcheng Huang}
\affiliation{Department of Chemistry,  Massachusetts Institute of Technology, Cambridge, MA 02139, USA}

\author[0000-0002-1921-4848]{Janusz J. Petkowski}
\affiliation{Department of Earth, Atmospheric and Planetary Sciences, Massachusetts Institute of Technology, Cambridge, MA 02139, USA}

\author[0000-0002-5147-9053]{Sukrit Ranjan}
\affiliation{Center for Interdisciplinary Exploration and Research in Astrophysics, Northwestern University, Evanston, IL 60201, USA}

\begin{abstract}
The search for signs of life on other worlds has largely focused on terrestrial planets. Recent work, however, argues that life could exist in the atmospheres of temperate sub-Neptunes. Here, we evaluate the usefulness of carbon dioxide isotopologues as evidence of aerial life. Carbon isotopes are of particular interest as metabolic processes preferentially use the lighter $^{12}$C over $^{13}$C. In principle, the upcoming James Webb Space Telescope (JWST) will be able to spectrally resolve the $^{12}$C and $^{13}$C isotopologues of CO$_{2}$, but not CO and CH$_{4}$. We simulated observations of CO$_{2}$ isotopologues in the H$_{2}$-dominated atmospheres of our nearest ($< 40$ pc), temperate (equilibrium temperature of 250-350 K) sub-Neptunes with M dwarf host stars.  We find $^{13}$CO$_{2}$ and $^{12}$CO$_{2}$ distinguishable if the atmosphere is H$_{2}$-dominated with a few percentage points of CO$_{2}$ for the most idealized target with an Earth-like composition of the two most abundant isotopologues, $^{12}$CO$_{2}$ and $^{13}$CO$_{2}$. With a Neptune-like metallicity of 100$\times$ solar and a C/O of 0.55, we are unable to distinguish between $^{13}$CO$_{2}$ and $^{12}$CO$_{2}$ in the atmospheres of temperate sub-Neptunes. If atmospheric composition largely follows metallicity scaling, the concentration of CO$_{2}$ in a H$_{2}$-dominated atmosphere will be too low to distinguish CO$_{2}$ isotopologues with JWST. In contrast, at higher metallicities, there will be more CO$_{2}$, but the smaller atmospheric scale height makes the measurement impossible. Carbon dioxide isotopologues are unlikely to be useful biosignature gases for the JWST era. Instead, isotopologue measurements should be used to evaluate formation mechanisms of planets and exoplanetary systems.

\end{abstract}

%%%%%%%%%%%%%%%%%%%%%%%%%%%%%%%%%%%%%%%%%%%%%%%%%%%%%%%%%%%%%%%%%%%
\section{Introduction} \label{sec:intro} 

So far the search for signs of life on other worlds has focused on terrestrial planets, but recent work has explored whether life could originate and survive in the liquid water clouds of temperate sub-Neptune-sized planets \citep{Seager2021}. Sub-Neptunes are more readily observable than terrestrial planets due to their larger size and low mean molecular weight (MMW), H$_{2}$-dominated atmospheres \citep[e.g.,][]{Bean2021}. Passive microbial-like particles could persist aloft in regions with liquid water clouds for long enough to metabolize, reproduce, and spread before sinking to altitudes that may be too hot for life of any kind to survive \citep{Seager2021}. 
Isotope fractionation is thought to be among the strongest signs of life that theoretically can be remotely detected in a planet's atmosphere \citep{Neveu2018}. On Earth, carbon isotope fractionation has been used as evidence of early life as biotic deposits of carbon have a higher ratio of $^{12}$C/$^{13}$C than abiotic deposits \citep[e.g.,][]{Rothschild1989}. The key metabolic processes that fractionate carbon are photosynthesis, chemosynthesis, respiration, and decomposition of organic matter \citep[e.g.,][]{Mackenzie2006}. Isotopic compositions, however, can also be changed by geophysical or chemical processes (e.g., degassing of basaltic magma \citep{Mattey1991} and chemical exchange in the protoplanetary disk \citep{Woods2009}). 

The upcoming James Webb Space Telescope (JWST) \citep{Gardner2006}, launched on December 25, 2021, provides an opportunity to characterize exoplanets using isotopologues. JWST is predicted to be able to make measurements of the protosolar D/H ratio ($2\times 10^{-5}$) using the methane isotopologue, CH$_{3}$D, for nearby, cool brown dwarfs in just 2.5 hr of JWST observing time \citep{Morley2019}. In addition, oxygen isotopes could be used to look for evidence of ocean loss. Simulated observations suggest that JWST could observe oxygen isotopologues in water and carbon dioxide molecules in the atmospheres of terrestrial planets around the M dwarf star TRAPPIST-1. The oxygen isotopic ratio, $^{18}$O/$^{16}$O, can be used as a marker of ocean loss and atmospheric escape as the lighter $^{16}$O is preferentially removed from photodissociated water vapor. Simulated JWST observations show that we may be able to distinguish Venus-like fractionation ($\delta^{18}$O 100) potentially caused by massive ocean loss from Earth-like fractionation in as little as 4 transits for TRAPPIST-1b and 5 transits for TRAPPIST-1d \citep{Lincowski2019}. 

From the ground, high-resolution data can be used to detect isotopologues via direct imaging and spectroscopy. Recently, $^{13}$CO became the first carbon isotopologue to be measured in an exoplanet's atmosphere \citep{Zhang2021}. Using the SINFONI instrument on the Very Large Telescope (VLT), \cite{Zhang2021} observed the atmosphere of super-Jupiter TYC 8998-760-1 b and found a $^{12}$CO/$^{13}$CO ratio of $31^{+17}_{-10}$ (90\% confidence) in contrast to our solar system average ratio of approximately 89 \citep{Woods2009} and local interstellar medium value of about 68 \citep{Milam2005}. \cite{Zhang2021} attribute the significant $^{13}$C enrichment to the formation location of the planet past the CO snowline where isotopic ion-exchange reactions and isotope-selective photodissociation can lead to an enhancement in $^{13}$C in CO ice compared to CO gas. 

Additionally, simulations of high-resolution observations of $^{13}$C$^{16}$O, HDO, and CH$_{3}$D have been created based on models of the CRyogenic InfraRed Echelle Spectrograph Upgrade (CRIRES+) instrument on the Very Large Telescope (VLT) and the Mid-infrared ELT Imager and Spectrograph (METIS) instrument on the upcoming Extremely Large Telescope (ELT) \citep{MolliereSnellen2019}. $^{13}$C$^{16}$O was detectable in the dayside emission of hot Jupiters using the simulated VLT observation \citep{MolliereSnellen2019}. 40 m-class telescopes, like the ELT, might make detection of CH$_{3}$D possible for cool planets \citep{MolliereSnellen2019}. CH$_{3}$D might also be detectable in nearby super-Earths if they are irradiated or transiting \citep{MolliereSnellen2019}. However, HDO observations may be hampered by methane absorption, which occurs at similar wavelengths. The ELT will likely be necessary for observations of HDO unless the methane is quenched \citep{MolliereSnellen2019}.  

%%%%% Paper Roadmap
We extend past work to focus on assessing the usefulness of isotopologues as biosignature gases. Here, we quantify the detectability of $^{12}$C/$^{13}$C via CO$_{2}$ in the atmospheres of nearby sub-Neptunes. We choose sub-Neptunes as they are more readily observable than super-Earths due to their larger size and low MMW, H$_{2}$-dominated atmospheres. We choose carbon dioxide isotopologues as carbon is fractionated by life, and the spectral separation of $^{12}$CO$_{2}$ from $^{13}$CO$_{2}$ is theoretically large enough to be resolved with JWST. In Section \ref{sec:methods}, we explain how we modeled the detection of isotopologues using spectra generated with {\tt petitRADTRANS} \citep{Molliere2019prt} and a simulated JWST noise model using {\tt PandExo} \citep{Batalha2017b}. In Section \ref{sec:results}, we show our results and explain how we assessed detectability. In Section \ref{sec:discussion}, we expound on the significance of our findings and evaluate the usefulness of isotopologues as biosignature gases in exoplanetary atmospheres. Our framework is designed to assess if carbon fractionation is a detectable biosignature gas or if the technology needed will be out of reach for decades to come. Knowing what kinds of measurements we can and cannot make will help guide future mission development.

%%%%%%%%%%%%%%%%%%%%%%%%%%%%%%%%%%%%%%%%%%%%%%%%%%%%%%%%%%%%%%%%%%%%%%%%
\section{Methods} \label{sec:methods} 

We used {\tt petitRADTRANS} to generate simulated spectra. {\tt petitRADTRANS} \citep{Molliere2019prt} is a well validated Python package to generate transmission spectra that has been broadly accepted by the community. As input to  {\tt petitRADTRANS}, we calculated $^{13}$CO$_{2}$ isotopologues cross sections using {\tt ExoCross} \citep{Yurchenko2018} from input line list that are available publicly on the High-Resolution Transmission Molecular Absorption (HITRAN) database \citep{Gordon2017}. We use the temperature-pressure profile of temperate sub-Neptune K2-18b by \cite{Blain2021}. We calculate chemical abundances using the chemical equilibrium implementation from {\tt petitRADTRANS}. The chemical abundances of the dominant volatiles at 1 bar are used as input boundary conditions to our photochemical model of the atmosphere. 

Our photochemistry model \citep{Hu2012} is a simplified one-dimensional chemical transport model. By calculating the atmosphere’s steady-state chemical compositions, our model can simulate a variety of atmospheres, such as H$_{2}$-dominated reducing atmospheres and CO$_{2}$-dominated highly oxidizing atmospheres. The photochemistry model has been validated by simulating modern Earth’s and Mars’ atmospheres \citep{Hu2012}. For more detailed information and applications of our model see \cite{Hu2012,  Hu2013, Seager2013, SousaSilva2020, Zhan2021, Huang2021}.

Our fiducial sub-Neptune atmosphere for the photochemical model is an H$_{2}$-dominated, Neptune-like atmosphere ([M/H]=100, C/O=0.55) orbiting an M dwarf star. We adopt the mass and radius of K2-18 b to describe the planet and the M dwarf stellar radiation model of GJ 876 \citep{France2016, Loyd2016}. Our simulated atmosphere consists of two segments: a lower convective layer ($10^{5}$ Pa to $10^{-1}$ Pa) and an upper radiative layer ($< 10^{-1}$ Pa). (The two segments are unrelated to our 100-layer atmosphere model used to simulate the transmission spectrum.) The temperature-pressure profile of the convective layer comes from Figure 4 of \cite{Blain2021}. Since we do not consider heating in the upper atmosphere, we assume the temperature to be constant (isothermal) in the upper radiative layer. In our photochemistry model, the vertical mixing processes are parameterized by the atmosphere's eddy diffusion coefficient ($K_{zz}$). We use the free-convection and mixing-length theories \citep{Stone1976,Flasar1977,Gierasch1985,Visscher2011} to help us constrain the value of $K_{zz}$. However, the constraints are loose, and hence we tested the impact of using a range of values of K$_{zz}$,  $1\times10^{8}$ to  $1\times10^{10}$ \citep{Gierasch1985,Visscher2011,Hu2014}. We found that the  K$_{zz}$ did not change our column averaged mixing ratios. In our fiducial simulation run, we adopt a uniform K$_{zz}$ of $1\times10^{9}$ cm$^{2}$s$^{-1}$, a value also adopted by \citet{Moses2013, Miguel2013}, and \citet{Hu2014}. In our simulation, we include all the reactions mentioned in \citet{Hu2012}, except for HSO$_{2}$ thermal decay, high-temperature reactions, and reactions that involve more than two carbon atoms \citep{Hu2012}. Given our choice of the temperature-pressure profile, we omit NH$_{4}$SH and NH$_{3}$ aerosols since the atmosphere is too hot for them to condense \citep{Hu2021PhotochemistryExoplanets}. Furthermore, we do not consider rainout in our simulation as there is no ocean and the temperature at the bottom of our simulated atmosphere is such that any rain will inevitably evaporate.

Using the results of our photochemistry model, we simulate observations of the Near Infrared Spectrograph (NIRSpec) Instrument on JWST. NIRSpec covers 0.6 to 5.3 $\mu$m \citep{STScI}. Using the PRISM disperser, the entire bandpass can be measured at a resolution of R$\approx$100. For high-resolution observations at an R$\approx$2700, we simulate the G395H disperser over the range of 2.87 $\mu$m to 5.27 $\mu$m \citep{STScI}. 

We calculate the anticipated on-orbit noise for JWST using the Space Telescope Science Institute Exposure Time Calculator for Exoplanets, {\tt PandExo} \citep{Batalha2017a}. {\tt PandExo} performs throughput calculations using the Space Telescope Science Institute's (STScI) Pandeia. \cite{Schlawin2021} estimated the magnitude of the noise for NIRCam, which has similar systematic error to NIRSpec. They found that the pointing uncertainty and jitter will be less than 6 ppm; thermal instabilities in the optics will be less than 2 ppm; persistence will be less than 3 ppm; and detector temperature fluctuations will be about 3 ppm. Given the sources of uncertainties, we adopt a noise floor of 10 ppm for our model. To further configure our noise model, we set an 80\% saturation level for each pixel. We also set the in-transit time equal to the out-of-transit time. We use input stellar spectrum from {\tt PHOENIX} \citep{Husser2013} based on the magnitude, temperature, metallicity, and surface gravity of the host star. In addition, we define the properties of the planet and host star.

%%%%%%%%%%%%%%%%%%%%%%%%%%%%%%%%%%%%%%%%%%%%%%%%%%%%%%%%%%%%%%%%%%%%%%%%%%%%%%%%
\section{Results} \label{sec:results} 

We present exoplanet transmission spectra with and without $^{13}$CO$_{2}$ in order to quantify the detectability of carbon isotopic composition in the atmosphere of a sub-Neptune-sized exoplanet with JWST. When $^{13}$C is included, we assume Earth-like fractionation. 

\subsection{Detection of Carbon Isotopes in an Ideal Target}

\begin{figure*}[htp]
    \centering
    \includegraphics[width=\linewidth]{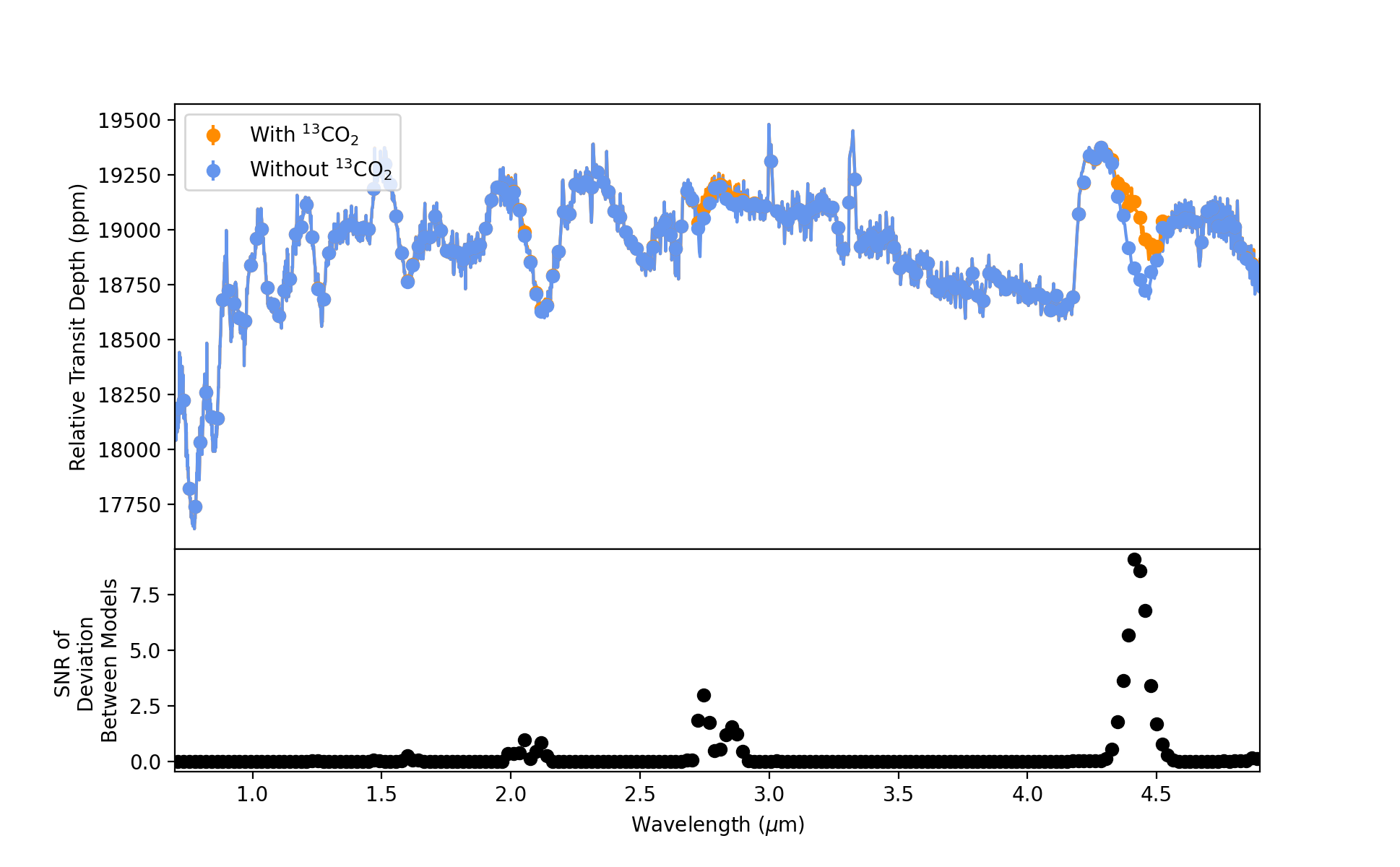}
    \caption{Simulated observation of the sub-Neptune TOI-1231 b artificially placed around our nearest M dwarf host star, LTT 1445 A, observed in the PRISM mode for 10 transits. The simulated atmosphere has an MMW of 2.42 and contains similar molecular abundances to those found in K2-18 b \citep{Benneke2019}. The upper panel shows the relative transit depth in parts per million (ppm) for the wavelengths associated with JWST's PRISM mode. The blue spectrum does not include $^{13}$CO$_{2}$ while the orange one does. The lower panel shows the signal to noise ratio (SNR) of the deviation between the models with and without $^{13}$CO$_{2}$. As seen in lower panel, for a 10 transit observation, the SNR value is 9.1.}
    \label{fig:toi1231benneke}
\end{figure*}

We find $^{13}$CO$_{2}$ to be detectable with JWST for the most idealized, simulated target. To best distinguish $^{13}$CO$_{2}$ from $^{12}$CO$_{2}$ the atmospheric composition needs to be predominately hydrogen gas with trace amounts of carbon dioxide. \cite{Benneke2019} found that K2-18 b has a H$_{2}$-dominated, a low MMW (2.42) atmosphere that contains water. While \cite{Benneke2019} did not conclusively detect other gases, they were able to place upper limits on CO, CO$_{2}$, NH$_{3}$, and CH$_{4}$ gas. We simulated an observation using a similar atmospheric composition based on the low MMW and upper limits. We found $^{13}$CO$_{2}$ was distinguishable from $^{12}$CO$_{2}$ at an SNR of 9.1 after a 10 transit observation as seen in Figure \ref{fig:toi1231benneke}. However, this result holds only for the most ideal artificial target of the sub-Neptune TOI-1231 b placed around our nearest M dwarf host star, LTT 1445 A. We are unable to distinguish $^{13}$CO$_{2}$ from $^{12}$CO$_{2}$ for any of the real temperate sub-Neptune targets.

\subsection{Non-Detection of Carbon Isotopes in Simulated Sub-Neptune Atmospheres}

\begin{figure*}[htp]
    \centering
    \includegraphics[width=\linewidth]{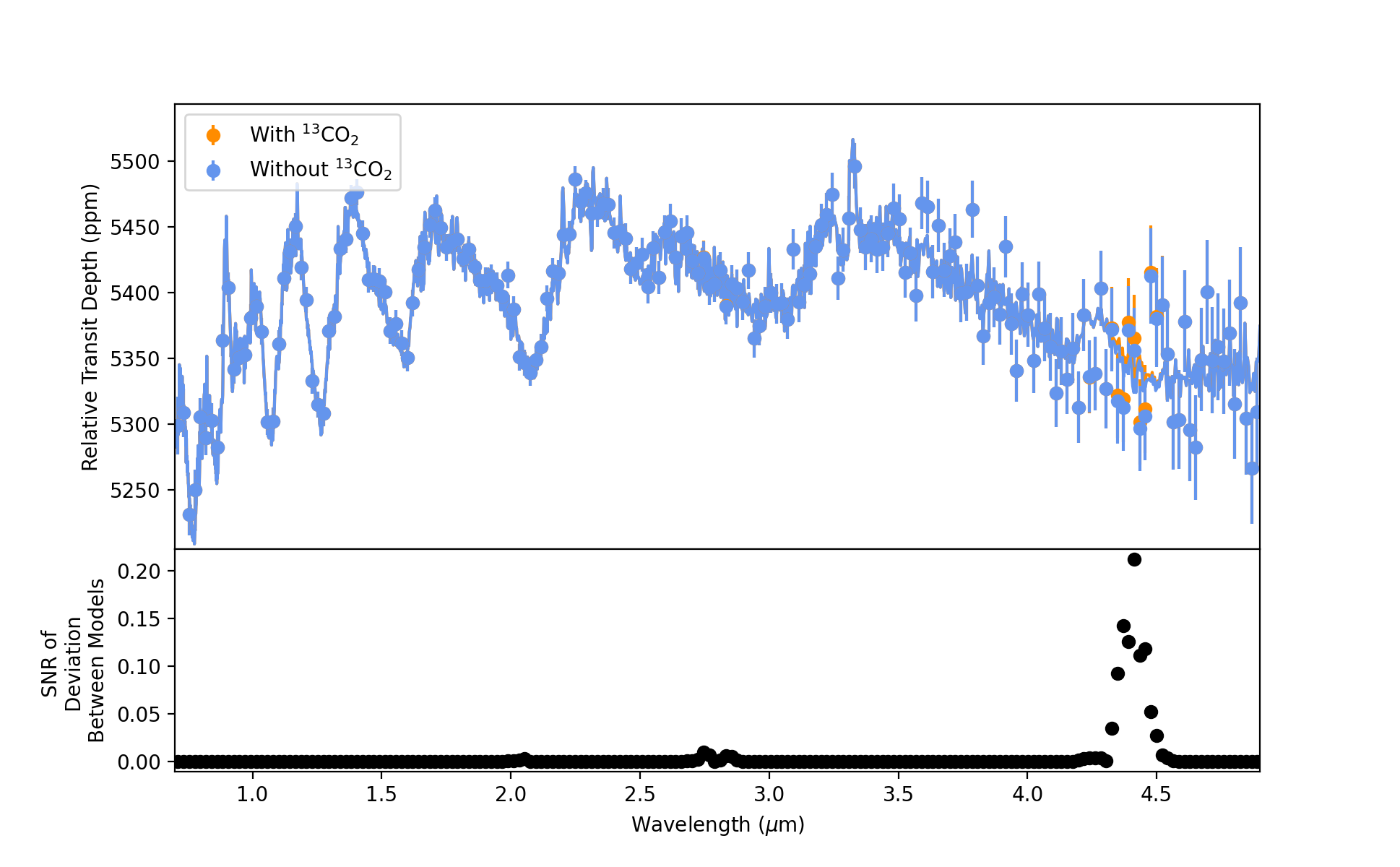}
    \caption{Simulated 10 transit, PRISM mode observation of TOI-1231 b with a Neptune-like metallicity of 100$\times$ solar and a C/O ratio of 0.55. The upper panel shows the relative transit depth in parts per million (ppm) for the wavelengths associated with JWST's PRISM mode. The blue spectrum does not include $^{13}$CO$_{2}$ while the orange one does. The lower panel shows the signal to noise ratio (SNR) of the deviation between the models with and without $^{13}$CO$_{2}$. As seen in lower panel, for a 10 transit observation, the SNR value is low at 0.2.}
    \label{fig:toi1231}
\end{figure*}

Our results show that we we will not be able to distinguish between carbon dioxide isotopologues for known temperate sub-Neptunes regardless of the atmospheric composition. We construct an atmosphere with a Neptune-like metallicity of 100x solar and a C/O ratio of 0.55 using our photochemistry model. We simulate a 10 transit observation for our most readily observable, temperate sub-Neptune, TOI-1231 b, with JWST's PRISM mode as shown in Figure \ref{fig:toi1231}. The upper panel shows the relative transit depth in parts per million (ppm) over the wavelength coverage of the PRISM mode (R $\sim$ 100). The blue spectrum includes only CO$_{2}$ with $^{12}$C while the orange spectrum also includes $^{13}$CO$_{2}$. The lower panel shows the signal-to-noise ratio of the deviation between the models or how well we would be able to distinguish observing CO$_{2}$ with and without $^{13}$C. The SNR for a 10 transit observation is very low at 0.2. 

\begin{figure*}[htp]
    \centering
    \includegraphics[width=\linewidth]{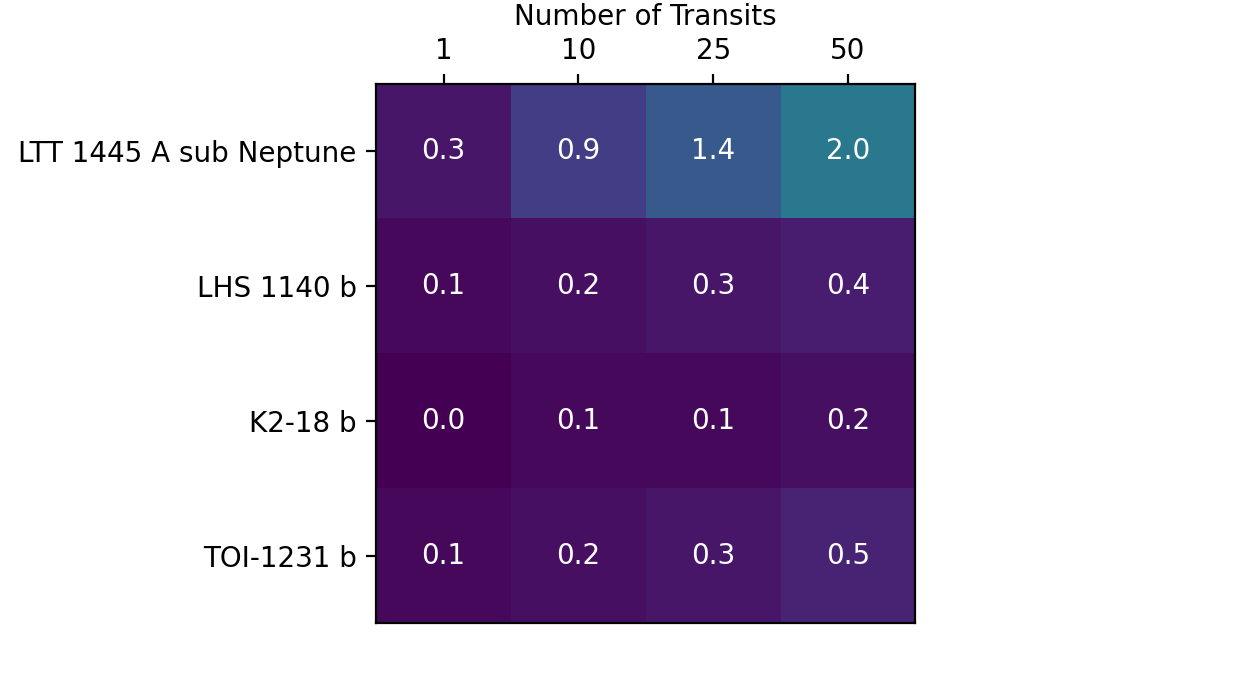}
    \caption{Summary plot of the SNR of the deviation between models that include $^{13}$C via CO$_{2}$ vs those that do not for an atmosphere with a Neptune-like metallicity of 100x solar and C/O of 0.55. The row labels show the names of the simulated target planets. The first planet is our artificial sub-Neptune (TOI-1231 b) around our closest known M dwarf host star LTT 1445 A, which in reality is host to a super-Earth, not a sub-Neptune. The column labels show the numbers of transits observed, from 1 to 50. A 10 transit observation is roughly the longest reasonable observation with JWST. The colors of the boxes correspond to the SNR value shown in each box where low values are cool colors with purple representing the lowest values. The SNR values range from 0.0 to 2.0, indicating that $^{13}$CO$_{2}$ is not detectable for a Neptune-like 100x solar metallicity atmosphere for even the most ideal, artificial sub-Neptune.}
    \label{fig:photochemzmax}
\end{figure*}

We also simulated observations of LHS 1440 b, K2-18 b, and an artificial sub-Neptune around our closest M dwarf host star, LTT 1445 A. In all cases, we found we were unable to distinguish between our models with and without $^{13}$CO$_{2}$. The SNR of the deviation between the model with and without $^{13}$CO$_{2}$ for the four different planets is shown in Figure \ref{fig:photochemzmax}. The planets are listed in the rows, and the number of transits observed changes with the columns. The first planet listed is an artificial planet, a sub-Neptune around the closest known M dwarf host star LTT 1445 A. (In reality, LTT 1445 A b is a 1.38 R$_{Earth}$ super-Earth, not a sub-Neptune.) 

We also tested the impact of varying the metallicity of the atmosphere and found that there is no ``sweet spot" between a low MMW atmosphere and a CO$_{2}$ abundance that allowed for a positive detection of $^{13}$CO$_{2}$ if the atmosphere followed metallicity scaling. The range of possible sub-Neptune atmospheric compositions remains largely unknown \citep[][and references therein]{Bean2021}. The compositions are believed to be H$_{2}$ dominated, though the metallicity range remains uncertain. While a lower metallicity atmosphere has a larger scale height, making it more readily observable, it comes at the cost of decreased CO$_{2}$ gas. A more metal-rich atmosphere would increase the amount of CO$_{2}$, but would also increase the MMW of the atmosphere, thus shrinking the scale height of the atmosphere and making it harder to detect overall, mostly due to large amounts of the heavy molecules, water and methane. Fortunately, CO$_{2}$ has a relatively large cross section, making it detectable at relatively low concentrations. In addition, carbon fractionation between the most abundant isotopologue and second must abundant isotopologue is relatively large at Earth-like fractionation. The carbon isotopologues of CO$_{2}$ are well separated compared with those of methane and carbon monoxide, allowing them to be potentially distinguishable even at low (R$\sim100$) spectral resolution such as with JWST. 

Currently, there are only four known nearby, temperate sub-Neptunes. LHS 1140 b is the closest at 12.47 $\pm$ 0.42 pc, but it is small for a sub-Neptune, right at the accepted cutoff between super-Earth and sub-Neptune at 1.727 $\pm$ 0.032 R$_{E}$ \citep{Dittmann2017, Ment2019}. It has an estimated equilibrium temperature of 235.0 $\pm$ 5 K \citep{Ment2019}. TOI-270 d is the next closest at 22.453 $\pm$ 0.021 pc and it has an estimated equilibrium temperature of 340 $\pm$ 14 K \citep{Gunther2019}. Farther away, TOI-1231 b is located at 27.4932 $\pm$ 0.0123 pc and has an equilibrium temperature of 330.0 $\pm$ 3.8 K \citep{Burt2021}. Lastly, K2-18 bis located at 38.07 $\pm$ 0.08 pc and has an estimated equilibrium temperature of 284 $\pm$ 15 K \citep{Sarkis2018}. 

As more TESS Targets of Interest (TOIs) become validated planets, it is likely that additional temperature sub-Neptunes around nearby, bright M dwarfs, emerge as better targets. However, additional temperate sub-Neptunes will not prove amenable to isotopologue detection with JWST if the atmospheres follow solar metallicity scaling.  

\subsection{Non-Detectability at Higher Spectral Resolution}
We also simulated observations using JWST's high-resolution G395H disperser. The simulations revealed that our inability to distinguish $^{13}$CO$_{2}$ from  $^{12}$CO$_{2}$ is not limited by the PRISM modes' lower resolution, but by the number of photons received during the observation. In the high-resolution mode, the number of photons per bin decreased, leading to increased noise that obscured the spectral differences between $^{13}$CO$_{2}$ and $^{12}$CO$_{2}$.

\subsection{Evaluation of Ground-based Observations}
We also considered high-resolution ground-based observations of our targets. The velocity of our proposed targets is not a limiting issue. However, the cross-correlation technique will not be possible for temperate sub-Neptunes as the contrast between planet and star is too small for current giant telescopes. The SNR in the photon limited regime can be calculated as $$\text{SNR}_{\text{planet}}=\frac{\text{S}_{\text{p}}}{\text{S}_{*}}\text{SNR}_{\text{star}}\sqrt{\text{N}_{\text{lines}}}$$ \citep{Birkby2018}. A coronagraph on a large ground-based telescope may allow for low-contrast observations in the future. Currently though, JWST is the best near-future instrument for measuring carbon dioxide isotopologues in the atmospheres of temperate planets.

%%%%%%%%%%%%%%%%%%%%%%%%%%%%%%%%%%%%%%%%%%%%%%%%%%%%%%%%%%%%%%%%%%%%%%%%
\section{Discussion} \label{sec:discussion}
Even if someday we can detect carbon fractionation via $^{13}$CO$_{2}$, it will be nearly impossible to prove that it is caused by life and not by another process. Today, we do not even have a clear understanding of the composition of sub-Neptune atmospheres. Much of the early observations with JWST will simply focus on first assessing if small planets have atmospheres or not and then on measuring their atmospheric composition. 

\subsection{The Challenge of Disentangling Abiotic and Biotic Isotope Fractionation}
A major complication of using isotopologues as biosignature gases is disentangling abiotic and biotic isotope fractionation. While isotope fractionation could be a strong indicator of life that is able to be detected remotely with the right instrument, we need to understand the context to rule out false positive detections \citep{Neveu2018}. The context includes both a habitability assessment of a planet and our understanding of abiotic fractionation processes. The magnitude of the fractionation alone will not be enough to unambiguously disentangle different biotic and abiotic fractionation processes.

On Earth, isotopic fractionation is measured by comparing the isotopic composition of a biotic carbon reservoir against an inorganic carbon reservoir. The composition of one isotope compared to another is not a biosignature on its own--it must be compared to a baseline inorganic value.

\subsubsection{Abiotic Carbon Fractionation}

Carbon fractionation occurs abiotically through processes such as chemical exchange reactions, photodissociation, gas release, and the kinetic isotope effect \citep{Quay1991,Woods2009}. One known planetary process that fractionates carbon is volcanism \citep{Mattey1991}. The degassing of basaltic magma enriches the melt in $^{12}$CO$_{2}$, while enhancing the gas in $^{13}$CO$_{2}$ \citep{Mattey1991}. The magnitude of this effect is only $\sim2$\textperthousand \citep{Mattey1991}. Our only example of a habitable planet (Earth) is subject to volcanic activity. Thus, carbon fractionation through magma outgassing will likely prove an important false positive that will need to be disentangled from biotic fractionation. 

Another possible cause of fractionation could be differences in bulk isotopic composition between the remote exosystem and our own. If for instance, the solar system formed in an area near intermediate-mass AGB stars, it would be enriched in $^{13}$C \citep{Kobayashi2011}. Understanding the composition of the baseline abiotic carbon reservoir is the key to being able to interpret atmospheric measurements of $^{13}$CO$_{2}$. 

\subsubsection{Using Carbon Fractionation via Methane to Rule Out False Positives}

Detection of carbon fractionation in an exoplanetary atmosphere can only be considered a sign of life if we have a full understanding of its context.  While biologically mediated fractionation of carbon dioxide through photosynthesis is too subtle to be detectable, other carbon-bearing species like CH$_{4}$ or CO may prove more amenable for fractionation studies. 

In contrast to our work assessing biologically fractionated carbon via CO$_{2}$, \cite{MolliereSnellen2019} suggested that metabolic carbon fractionation could be measured by comparing the composition of $^{13}$C vs $^{12}$C in CO$_{2}$ vs CH$_{4}$. Isotopic composition via CO$_{2}$ would serve as the baseline inorganic carbon reservoir while isotopic composition via CH$_{4}$ would be considered the organic carbon reservoir. Thus, an important application of our framework could be to assess the detectability of carbon fractionation via methane. However, as the spectral resolution required is too high for JWST and the contrast between host star and temperate planet is too low for ground-based high-resolution observations, the path forward to measure biologically driven isotope fractionation seems bleak.

\subsection{Metabolic Carbon Fractionation Pathways}

Several metabolic processes are known to fractionate carbon, releasing waste gases like CO$_{2}$ and CH$_{4}$ that can accumulate in the atmosphere. In the context of a H$_{2}$-dominated sub-Neptune, there are three potentially key metabolic strategies that are capable of altering the atmospheric carbon isotopic ratio: (1) hydrogenic photosynthesis, (2) methanogenesis, and (3) anaerobic oxidation of methane or organic matter. 

\subsubsection{Hydrogenic Photosynthesis}
Photosynthesis, or the conversion of light energy to chemical energy, is one of the oldest and most fundamental energy-harnessing biochemical processes on Earth. Photosynthesis is broadly accepted as likely to occur on other planets with life \citep[e.g.,][]{Seager2005,Kiang2007}. Earlier work has evaluated the detectability of the vegetation ``red edge" and the byproducts of photosynthesis, such as O$_{2}$, along with its secondary products \citep{Seager2005, Kiang2007}. Even if photosynthesis itself might be a cosmically common phenomenon, life on other worlds might use different kinds of photosynthesis than life on Earth. On Earth, oxygenic photosynthesis fractionates CO$_{2}$, but in a H$_{2}-$dominated sub-Neptune atmosphere, hydrogenic photosynthesis is the more likely form \citep{Bains2014}.

Hydrogenic photosynthesis uses CH$_{4}$ instead of CO$_{2}$ as a carbon source. Thus, hydrogenic photosynthesis does not directly contribute to the atmospheric CO$_{2}$ isotopic fractionation. However, it can lead to fractionation of CH$_{4}$ and the carbon isotopic fractionation of the produced organic matter (CH$_{2}$O) through the following reaction: CH$_{4}$ + H$_{2}$O $\rightarrow$ CH$_{2}$O + 2H$_{2}$. The substrates of hydrogenic photosynthesis, CH$_{4}$ and H$_{2}$O, are readily available in the atmospheres of sub-Neptunes as CH$_{4}$ is expected to be the dominant form of carbon and H$_{2}$O the dominant form of oxygen. CO$_{2}$, on the other hand, is likely rare and as a result an unlikely substrate for photosynthesis. The product of the hydrogenic photosynthesis, just like in oxygenic photosynthesis, is organic matter (CH$_{2}$O). Life preferentially uses the lighter $^{12}$C in reactions, leading to a preferential accumulation of $^{12}$CH$_{2}$O. Hydrogenic photosynthesis could therefore indirectly lead to CO$_{2}$ fractionation in sub-Neptune atmospheres as the anaerobic oxidation of the produced $^{12}$CH$_{2}$O organic matter could in principle result in the release of more $^{12}$CO$_{2}$. 

\subsubsection{Methanogenesis}
Methanogenesis is another process known to cause carbon fractionation. When organisms perform methanogenesis, they catalyze the reduction of CO$_{2}$ to CH$_{4}$ and release energy through the following reaction: CO$_{2}$ + 4H$_{2} \rightarrow $ CH$_{4}$ + 2H$_{2}$O. The reduction of CO$_{2}$ by H$_{2}$ could be a ubiquitous source of energy for life in a H$_{2}$-rich atmosphere \citep{Seager2021}. Methanogenesis can proceed even under conditions where only trace amounts of CO$_{2}$ are available, such as in the H$_{2}$-dominated atmospheres of sub-Neptunes \citep{Bains2014}. Therefore, life can yield enough energy released by the reduction of CO$_{2}$ to CH$_{4}$ with small amounts of CO$_{2}$ being regenerated by photochemistry. If life preferentially acquires $^{12}$CO$_{2}$ from the atmosphere as a substrate for methanogenesis, then a potential for biologically-driven CO$_{2}$ isotopic fractionation exists as more $^{13}$CO$_{2}$ will be left in the atmosphere as more $^{12}$CH$_{4}$ is produced.

\subsubsection{Anaerobic Oxidation of Methane}
The third possible metabolic strategy for life in a sub-Neptune aerial biosphere is anaerobic oxidation of methane (AOM) or anaerobic oxidation of organic matter \citep{Timmers2017}. The AOM process is a type of respiration that oxidizes CH$_{4}$ with sulfate, nitrate, nitrite, or metal oxides for example: $$CH_{4} + 4 NO_{3}^{-} \rightarrow CO_{2} + 4 NO_{2}^{-} + 2 H_{2}O$$ \citep{Haroon2013}. Methane and organic matter oxidation reactions are possible in sub-Neptune atmosphere conditions where there is an abundance of CH$_{4}$. The potential challenges for AOM reactions in sub-Neptunes involve insufficient abundance of non-volatile oxidants like sulfates, nitrates, nitrites, or metals \citep{Seager2021}. AOM might be the source of biologically fractionated CO$_{2}$ where the preferentially released CO$_{2}$ is the lighter $^{12}$CO$_{2}$ variant, if life efficiently acquires the light $^{12}$CH$_{4}$ substrate for AOM. 

\subsection{Carbon Isotopic Composition Measurements with Future Mission Concepts}
We have also considered the feasibility of detecting CO$_{2}$ isotopologues with future space-based mission concepts. In order to distinguish CO$_{2}$ isotopologues in the atmospheres of sub-Neptunes, we need telescopes with apertures larger than JWST. We estimate the required size of the telescopes using a simple scaling relationship. The SNR is equal to the number of photons received on the detector divided by the noise. In the best case scenario, the noise is limited by the shot noise from the target star and is equal to the square root of the number of photons. As the number of photons scales proportionally to the square of a telescope's diameter, the SNR scales linearly with telescope aperture (SNR $\propto$ D). With our fiducial atmospheric composition of 100$\times$ solar and C/O=0.55, it will take a 20 m-class telescope to distinguish carbon dioxide isotopologues in 25 transits even for a sub-Neptune artificially placed around our closest M dwarf host star. If we instead consider a low MMW atmosphere with similar molecular abundances to the upper limits measured in K2-18 b  \citep{Benneke2019}, it would take only a 10 m-class telescope to distinguish the two most abundant carbon dioxide isotopologues in the atmosphere of TOI-1231 b. As of yet, there is no upcoming space-based telescope of this size. However, the next generation giant ground-based telescopes will be in the 30 to 40 m range.

\subsection{The Future of Carbon Fractionation as a Biosignature Gas}

The future of using carbon fractionation to assess signs of life on remote worlds is not bright. Most importantly, baseline carbon fractionation values for other exoplanetary systems are not known. In the solar system, carbon fractionation is largely homogenized \citep{Woods2010}. Notably, despite Earth being the only planet known to harbor photosynthetic life, its bulk carbon fractionation does not stand out from the rest of the solar system. In addition, planed and near-future instruments will be unable to measure subtle changes to atmospheric fractionation caused by metabolic processes. Isotopologue measurements for the next several decades should focus on evaluating planetary processes rather than biologically mediated fractionation. There are other more readily detectable biosignature gases for missions like the JWST to focus on such as O$_{2}$, O$_{3}$, CH$_{4}$, N$_{2}$O, CH$_{3}$Cl, and sulfur gases \citep{Schwieterman2018}.

The search for life in gaseous exoplanet atmospheres is particularly compelling because it offers the possibility of empirically testing environmental requirements for the the origin of life (abiogenesis) and, in particular, whether rocky planetary surfaces are required for abiogenesis. While origin-of-life theories invoking droplets have been proposed, their experimental study is limited and debated, and they generally invoke spray originating from terrestrial surface interfaces \citep{Donaldson2004OrganicHypothesis, Nam2017, Jacobs2019StudyingIonization}. Instead, most current theories of the emergence of life on Earth invoke the presence of a surface. Surfaces are invoked in prebiotic chemistry as a source of mineral reagents, catalysts, or scaffolds \citep[e.g.,][]{Corliss1981, Ferris2005MineralRNA, Kim2016, Adam2018}, redox disequilibrium through connection of a reducing interior to an oxidizing fluid environment \citep{Barge2017}, microenvironments that produce favorable local conditions for prebiotic chemistry not accessed in global mean conditions \citep[e.g.,][]{Ranjan2019NitrogenEarth, Toner2020ALife}, and heterogenous reservoirs whose coupling can enable chemistry not possible in a one-reservoir scenario (multi-pot syntheses; e.g., \citealt{Ritson2018}). Detection of life on a gaseous exoplanet atmosphere would require extremely stringent evidence that could outweigh, e.g., the prior belief in their uninhabitability on the basis of experimentally motivated theories of the origin of life to date \citep{Catling2018ExoplanetAssessment, Seager2021}. If such evidence could be found, however, it would empirically demonstrate the existence of a pathway to abiogenesis without surfaces, opening new vistas in the study of prebiotic chemistry, so long as panspermic transfer from a neighboring rocky body could be ruled out \citep{Lingam2017EnhancedSystem, Chen2018OnPerspective, Rimmer2021LifesExoplanets}. Carbon fractionation of atmospheric CO$_{2}$, while promising in principle, is unfortunately practically unlikely to provide such evidence; the search for robust gaseous exoplanet biosignatures continues.

%%%%%%%%%%%%%%%%%%%%%%%%%%%%%%%%%%%%%%%%%%%%%%%%%%%%%%%%%%%%%%%%%%%%%%%%%%%%%%%%
\section{Conclusion} \label{sec:conclusion} 

It is essential to evaluate which biosignature gases will prove the most fruitful for future observations of exoplanet atmospheres and which will never be detectable. We have developed a robust framework to test which, if any, isotopologues will be detectable biosignature gases for signs of remote life. Here, we applied our framework to carbon dioxide isotopologues to assess if $^{12}$CO$_{2}$ and $^{13}$CO$_{2}$ at Earth-like abundances could be distinguishable in the atmospheres of temperate sub-Neptunes orbiting M dwarf stars with the upcoming JWST. We selected carbon isotopes as metabolic processes preferentially use $^{12}$C over $^{13}$C. However, biological processes like photosynthesis drive subtle changes to the carbon fractionation ratio. As a first step, we assessed if we could distinguish between $^{12}$CO$_{2}$ and $^{13}$CO$_{2}$ at Earth-like abundances. 

We found that we will be able to distinguish the presence of $^{13}$C vs $^{12}$C via CO$_{2}$ at Earth-like abundances in the atmospheres of temperate sub-Neptunes with H$_{2}$-dominated atmospheres using JWST, but only for the most favorable atmospheric compositions and idealized targets. Thus, it is unlikely that  $^{13}$CO$_{2}$ will be measured in the atmosphere of a temperate sub-Neptune with JWST. Furthermore, ground-based high spectral resolution instruments will not enable isotopic detection in temperate sub-Neptune atmospheres due to the low contrast between the M dwarf host star and cool planet. When we are able to measure carbon fractionation in exoplanet atmospheres, our understanding of the system's context will be essential to prove if the observed fractionation is a possible sign of life. Isotopologue measurements for the next decade should focus on disentangling the physical processes that drive carbon fractionation rather than trying to detect them as unambiguous biosignature gases. 
\\~\\

%\acknowledgement 
We would like to acknowledge Dr. Kaitlin Rasmussen for a productive conversation about the feasibility of ground-based spectroscopy of CO$_{2}$ isotopologues in the atmospheres of temperate sub-Neptunes.

\bibliography{references.bib}{}
\bibliographystyle{aasjournal}

\end{document}